\documentclass[epj,referee,onecolumn]{svjour}
\usepackage{graphics}
\usepackage{amssymb}
\usepackage{color}

\newcommand{\op}[1]{%
    \fontdimen12\textfont3=2pt\fontdimen12\scriptfont3=1.4pt%
    \!\null\mathop{\vphantom{#1}\smash{#1}}\limits_{\sim}\null\!}

\newcommand{\Vektor}[3]{ \left(
\begin{array}{c}#1\\#2 \\#3 \end{array}\right) }
\spnewtheorem{ass}{Assumption}{\bf}{\it}

\begin{document}
\title{Exact ground state of a frustrated integer-spin modified Shastry-Sutherland model}
%\subtitle{Do you have a subtitle?\\ If so, write it here}
\author{Johannes Richter\inst{1} \and Heinz-J\"urgen
Schmidt\inst{2}% etc
% \thanks is optional - remove next line if not needed
%\thanks{\emph{Present address:} Insert the address here if needed}%
}                     % Do not remove
%
%\offprints{}          % Insert a name or remove this line
%
\mail{\tt \\Johannes.Richter@Physik.Uni-Magdeburg.DE}
\institute{Institut f\"ur Theoretische Physik, Otto-von-Guericke-Universit\"at Magdeburg,\\
PF 4120, D - 39016 Magdeburg, Germany\\ \and Universit\"at
Osnabr\"uck, Fachbereich Physik, Barbarastr. 7, D - 49069
Osnabr\"uck, Germany}
\date{Received: date / Revised version: date}
% The correct dates will be entered by Springer
%
\abstract{ We consider a two-dimensional geometrically  frustrated
integer-spin Heisenberg system that admits an exact ground state.
The system corresponds to a decorated square lattice with two
coupling constants $J_1$ and $J_2$, and it can be understood as a
generalized  Shastry-Sutherland model. Main elements of the spin
model are suitably coupled antiferromagnetic spin trimers with
integer spin quantum numbers $s$ and their ground state $\Phi$
will be the product state of the local singlet ground states of
the trimers.
We provide exact numerical data for finite lattices as well as
analytical considerations to estimate the  range of the  existence
in dependence  on the ratio of the two couplings constants $J_2$
and $J_1$ and on the spin quantum number $s$. Moreover, we find
that the magnetization curves as a function of the applied
magnetic field shows  plateaus and jumps.
In the classical limit $s\to\infty$ the model exhibits
phases of three- and two-dimensional ground states separated by a
one-dimensional (collinear) plateau state at 1/3 of the saturation
magnetization.}
 %end of abstract
%
\authorrunning{J.~Richter and H.-J.~Schmidt}
\titlerunning{Exact ground state of a modified Shastry-Sutherland model}
\maketitle
\section{Introduction}
\label{sec I}
The concept of frustration  plays an important role
in the search for novel quantum states of condensed matter, see,
e.g., \cite{lnp04,buch2,moessner01,frust1,frust2}. The
investigation of frustrating quantum spin systems is a challenging
task. Exact statements about the properties of  quantum spin
system are known only in exceptional cases. The simplest known
exact eigenstate is the fully polarized ferromagnetic state.
Furthermore the one- and two-magnon excitations above the fully
polarized ferromagnetic state  also can be calculated exactly,
see, e.g., \cite{mattis81,Kuzian07,Zhitomirsky10,nishimoto2011}. An
example for non-trivial eigenstates is Bethe's famous solution for
the one-dimensional (1D) Heisenberg antiferromagnet (HAFM)
\cite{bethe}.
The investigation of strongly frustrated magnetic systems
surprisingly led to the discovery of several new exact
eigenstates. Some of the eigenstates found for  frustrated quantum
magnets are  of quite simple nature and for several physical
quantities, e.g., the spin correlation functions, analytical
expressions can be found. Hence such exact eigenstates may play an
important role either as groundstates of real quantum magnets or
at least as groundstates of idealized models which can be used as
reference states for more complex quantum spin systems.
A well-known class of exact eigenstates are dimerized singlet
states, where a direct product of pair singlet states is an
eigenstate of the quantum spin system.  Such states become
groundstates for certain values/regions of frustration. The most
prominent examples are the Majumdar-Gosh state of the 1D $J_1-J_2$
spin-half HAFM \cite{majumdar} and the orthogonal dimer state of
the Shastry-Sutherland  model, see, e.g.,
\cite{shastry81,Mila,Miyahara,Lauchli,uhrig2004,darradi2005}. Many other frustrated spin
models in one, two or three dimensions are known which have also
dimer-singlet product states as groundstates, see, e.g.,
\cite{pimpinelli,ivanov97,japaner3d,koga,schul02}. A systematic
investigation of systems with dimerized eigenstates can be found
in  \cite{schmidt05}. Note that these dimer-singlet product
groundstates  have gapped magnetic excitations and lead therefore
to a plateau in the magnetization at $m=0$.
Recently it has been demonstrated for the 1D counterpart of the
Shastry-Sutherland model \cite{ivanov97,koga,schul02}, that more
general product eigenstates containing chain fragments of finite
length can lead to an infinite series of magnetization plateaus
\cite{schul02}.

Other examples of product ground states are single-spin product
states of 1D XYZ model \cite{mueller85} and the the highly
degenerate ground-state manifold of localized-magnon states found
for antiferromagnetic quantum spin systems on various frustrated
lattices \cite{lm}. Finally, we mention the so-called central-spin
model or Heisenberg star where also exact statements on the
groundstate are known \cite{starI}.

Although, at first glance such singlet-product states seem to
exist only for 'exotic' lattice models, it turned out that such
models are not only a playground of theoreticians but may become
relevant for experimental research. The most prominent example is
the above mentioned Shastry-Sutherland model introduced in 1981
\cite{shastry81} for which only in 1999 the corresponding
quasi-two-dimen\-sional compound SrCu$_2$(BO$_3$)$_2$ was found
\cite{Kage,srcubo}. Other examples are the quasi-1D spin-Peierls
compound $CuGeO_3$, see, e.g., \cite{cugeo},  or the star-lattice
compound
[Fe$_3$($\mu_3$-O)($\mu$T-OAc)$_6$-(H$_2$O)$_3$][Fe$_3$($\mu_3$-O)($\mu$-OAc)$_{7.5}$]$_2
\cdot$7 H$_2$O.\cite{star-exp,star-theor}

In the present paper we combine the ideas of Shastry and
Sutherland \cite{shastry81} and our recent findings on exact
trimerized singlet product ground states (TSPGS's) for 1D
integer-spin Heisenberg systems \cite{schmidt10} and discuss such
TSPGS's on a  two-dimensional modified Shastry-Sutherland
square-lattice model. Section \ref{sec_egs} shortly recapitulates
the theory of TSPGS's and section \ref{sec_model} defines the
modified  Shastry-Sutherland model and its finite realizations
that will be analyzed in what follows. We have concentrated in our
numerical studies on the size of the gap for the exact ground
state for finite lattices of $N=12$ (for spin  quantum numbers
$s=1$, $s=2$), as well as  $N=18$ and $N=24$ (for $s=1$) and on
the magnetization curves for selected values of $J_2$, see section
\ref{sec_nr}. The analytical results in section \ref{sec ar}
mainly concern upper and lower bounds of the gap function. These
results depend on a slightly generalized statement and proof of
the gap theorem, first formulated in \cite{schmidt10}, which is
done in appendix~\ref{app}. Finally, appendix~\ref{class} contains exact results on
classical magnetization curves for the model under consideration.

\section{Exact ground states}
\label{sec_egs} The anti-ferromagnetic uniform spin trimer
\begin{equation}\label{egs1}
H_1=J(\op{\bf{s}}_0\cdot\op{\bf{s}}_1+
\op{\bf{s}}_0\cdot\op{\bf{s}}_2+\op{\bf{s}}_1\cdot\op{\bf{s}}_2)
\end{equation}
has, for $J>0$ and integer $s$, a unique $S=0$ ground state, denoted
by $[0,1,2]$, with ground state energy
\begin{equation}\label{egs2}
E_0=-\frac{3}{2}J s(s+1) \;.
\end{equation}
The corresponding product state
\begin{equation}\label{egs3}
\Phi=\bigotimes_{i=1}^{\mathcal N}[i0,i1,i2]
\end{equation}
will be an eigenstate of a system of ${\mathcal N}$ coupled spin
trimers indexed by $i=1,\ldots,{\mathcal N}$ with Hamiltonian
\begin{equation}\label{egs4}
H=\sum_{i\epsilon
j\delta}J_{i\epsilon,j\delta}\,\op{\bf{s}}_{i\delta}\cdot\op{\bf{s}}_{j\epsilon}
\;,
\end{equation}
if and only if the coupling between different trimers is ``balanced"
in the following sense:
\begin{equation}\label{egs5}
J_{i\delta,j\delta}+J_{i\epsilon,j\epsilon}=J_{i\delta,j\epsilon}+J_{i\epsilon,j\delta}
\end{equation}
for all $1\le i<j\le{\mathcal N}$ and $\delta,\epsilon=0,1,2$, see
\cite{schmidt10}. Moreover, (\ref{egs3}) will be a ground state of
(\ref{egs4}), a TSPGS, if the intra-trimer coupling is almost
uniform and the inter-trimer coupling is not too strong
\cite{schmidt10}.
The domain of coupling constants where this is the case will be called the ``TSPGS-region".\\

If the system of trimers has a periodic lattice structure, the
difference $\Delta E$ between the energy of the first excited state
and that of the ground state can be shown \cite{schmidt10} to be
bounded from below independently of the system size. In other
words, the TSPGS is ``gapped".

\begin{figure}
\resizebox{0.9\columnwidth}{!}{
\includegraphics{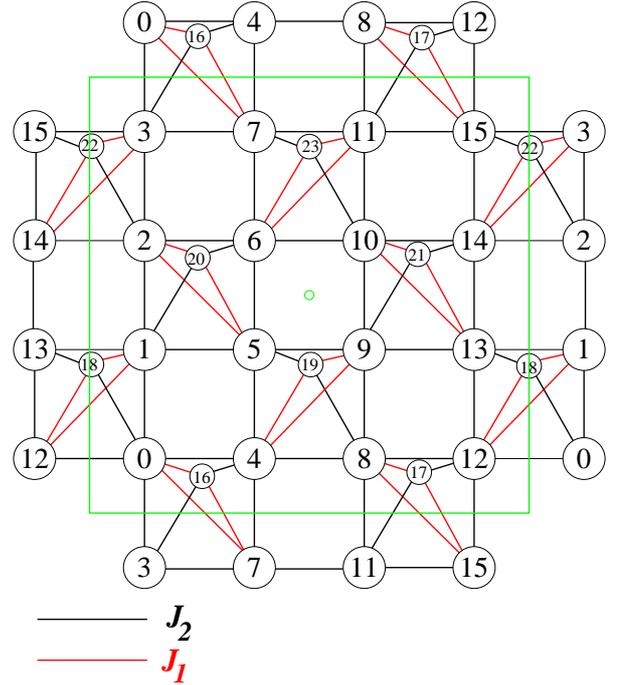}
} \caption{ The modified Shastry-Sutherland model on the decorated
square lattice for  $N= 24$ sites (periodic conditions imposed) used for exact diagonalization.}
\label{fig1}
\end{figure}

\begin{figure}
\resizebox{1.0\columnwidth}{!}{
\includegraphics{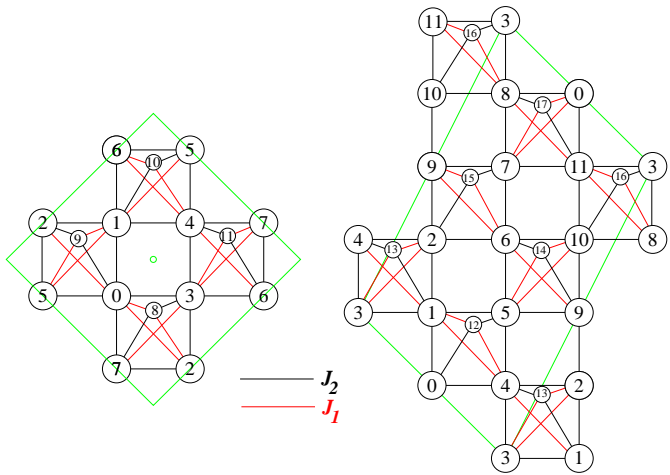}
} \caption{Two finite decorated square lattices of  $N= 12$  and
$N=18$ sites used for exact diagonalization.} \label{fig2}
\end{figure}

\section{The model}
\label{sec_model}

We consider the inter-spin Heisenberg Hamiltonian on a decorated
square-lattice (see figure~\ref{fig1}). It results from the
well-known Shastry-Sutherland model by replacing its diagonals by
equilateral triangles with uniform intra-trimer interaction strength
$J_1>0$. The set of triangles is divided in a bi-partite fashion
into two disjoint subsets of triangles of type I and type II,
corresponding to diagonals with positive slope resp.~negative ones,
see figure~\ref{fig1}. Each triangle of, say, type I is surrounded
by four
triangles of type II and connected to each of them with three bonds of strength $J_2$.\\
It follows that the inter-trimer coupling satisfies the balance
condition (\ref{egs5}) and hence the theory of TSPGS's applies. In
particular, two questions arise which will be addressed in the
following sections: What is the size of the TSPGS-region and of what
kind are the lowest excitations? The latter question is also
connected to the issue of magnetization plateaus which will be
shortly discussed below.

\section{Results}

\subsection{Numerical results}
\label{sec_nr} In what follows we set $J_1=1$ and consider $J_2$ as
the variable bond strength. To study the region where the TSPGS is
the ground state of the model (\ref{egs4}) we use the Lanczos exact
diagonalization (ED) technique. Since for spin quantum numbers $s >
1/2$ considered here the size of the Hamiltonian matrix grows much
faster with system size $N$ than for $s=1/2$, we are restricted to
finite lattices of $N= 12,18$ and $24$ for $s=1$ and $N=12$ for
$s=2$. The largest lattice is shown in figure~\ref{fig1}, whereas
the smaller lattices are shown in figure~\ref{fig2}. Although the
criterion for the existence of TSPGS's (see section \ref{sec_model})
are fulfilled, we have to mention that for the small lattices of
$N=12$ and $N=18$ the exchange pattern of the $J_1$ diagonal bonds
in the squares do not match to the infinite system. Nevertheless, we
have included the data for $N= 12$ and $18$ to get an impression on
finite-size effects and on the influence of the spin quantum number
$s$.

\begin{figure}
\vspace*{5cm}
\scalebox{0.7}{\includegraphics{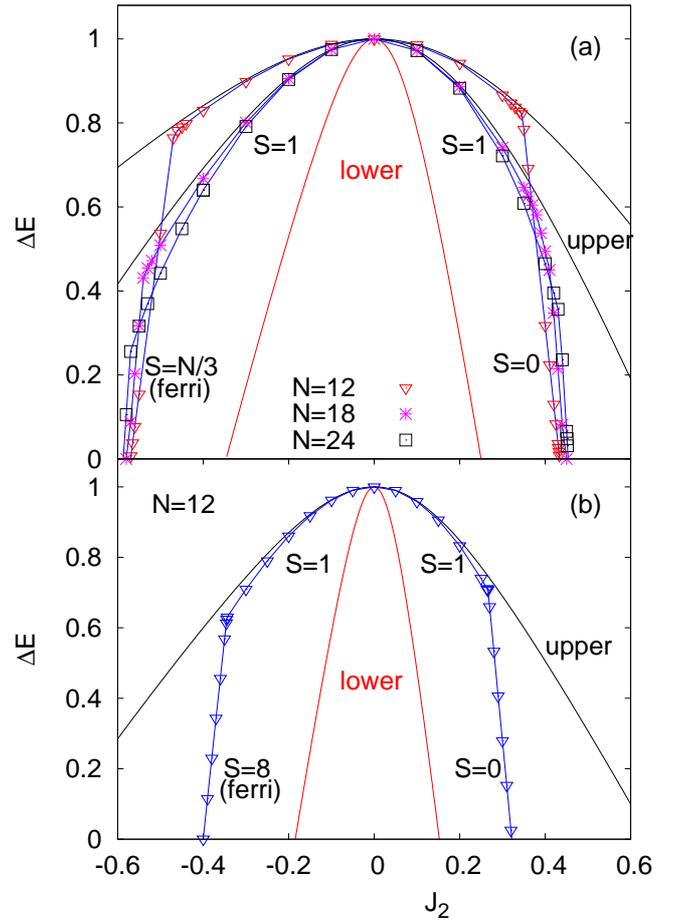}}
\caption{Numerical exact data for $N=12$, $18$, and
$24$ (symbols) as well as upper (black solid line) and lower bounds
(red solid line) for the excitation gap $\Delta E $. (a) spin
quantum number $s=1$;
(b) spin quantum number $s=2$.
Note that the labels $S=1$, $S=0$, $S=2N/3$ (ferri), and $S=8$
(ferri) characterize the total spin of the  excited state.
}
\label{fig3}       % Give a unique label
\end{figure}

According to \cite{schmidt10} the TSPGS is gapped. Hence we use the
spin gap, see figure~\ref{fig3}, to detect the critical points
$J^{c1}_2$ and $J^{c2}_2$, where the  TSPGS gives way for other
ground states. We find for $s=1$ the values $J^{c1}_2
=-0.570,-0.578$, and $-0.587$ and  $J^{c1}_2 =0.434, 0.446$, and
$0.454 $  for $N=12,18$, and $24$, respectively (cf.
figure~\ref{fig3}(a)). For $s=2$ and $N=12$ we have $J^{c1}_2
=-0.400$ and  $J^{c2}_2 =0.322$, cf. figure~\ref{fig3}(b). These
values lie between the upper and lower bounds which will be derived
for $J^{c1}_2$ and  $J^{c2}_2$ in the next section for $N \to
\infty$. The nature of the lowest excited state depends on $J_2$.
Around  $J_2=0$ it is a triplet state with strong antiferromagnetic
correlations along the trimer bonds and weak correlations between
the trimers. Near $J^{c1}_2$ the lowest excitation is a
ferrimagnetic state, i.e. the total spin is $S=Ns/3$ and the system
splits into two ferromagnetically correlated sublattices containing
on the one hand the $2N/3$ square-lattice sites (i.e. sites
$0,1,\ldots,15$ in figure~\ref{fig1}) and on the other hand the
$N/3$ additional sites (i.e. sites $16,17,\ldots,23$ in
figure~\ref{fig1}). The spin correlations between both sublattices
are anti-ferromagnetic. The ferrimagnetic state is the ground state
for $-1.5 < J_2 < J^{c1}_2$. Near $J^{c2}_2$ the lowest excitation
is a collective singlet state with strong correlations along all
bonds, and, this state becomes
the ground state at $J_2=J^{c2}_2$.

\begin{figure}
\vspace*{5cm}
\scalebox{0.7}{\includegraphics{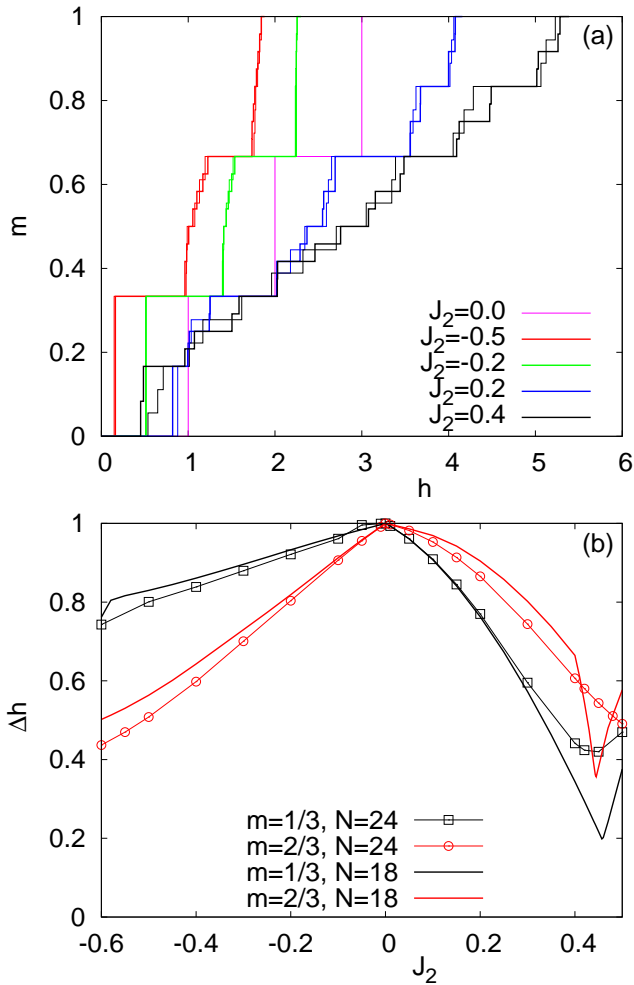}}
\caption{(a) Magnetization curve $m(h)$ for selected
values of $J_2$ and $s=1$ (thick lines $N=24$, thin lines $N=18$);
(b) Plateau widths $\Delta h$  of the $m=1/3$ and the  $m=2/3$ plateaus  as
a function of $J_2$ for $N=24$ and $N=18$ and $s=1$.
}
\label{fig4}       % Give a unique label
\end{figure}

It is well  known that the magnetization curve of the
Shastry-Sutherland model (as well as that of the corresponding
material SrCu$_2$(BO$_3$)$_2$) possesses a series of pla\-teaus, see,
e.g., \cite{Kage,kodama,misguich,mila}. Motivated by this, we study
now briefly the  magnetization curve $M(h)$ (where $M$ is the total
magnetization  and  $h$ is the strength of the external magnetic
field) for the considered model for $s=1$ using ED
for $N=18$ and $N=24$ sites. ED results  for the
relative magnetization $m=M/M_{sat}$ versus magnetic field $h$ for
$N=18$ and $N=24$ sites are shown in figure~\ref{fig4}a. Again the
finite-size effects seem to be small. Trivially, in the limit
$J_2=0$ the $m(h)$ curve consists of three equidistant plateaus and
jumps according to the magnetization  curve of an individual
triangle. Switching on a ferromagnetic inter-triangle bond $J_2 < 0$
the general shape of the magnetization curve is preserved. However,
the  saturation field as well as the end points of the plateaus
decrease almost lineraly with $J_2$ and become zero at $J_2=1.5$,
where the ground state becomes  the fully polarized ferromagnetic
state.

In case of a moderate antiferromagnetic inter-triangle bond $J_2 >
0$ the plateaus at  $m=1/3$ and $m=2/3$ still exist, however the
discontinuous transition between plateaus becomes smooth. Note that
a $m=1/3$ plateau was also found for the standard Shastry-Sutherland
model \cite{misguich,mila}. The plateau widths $\Delta h$ of the $m=1/3$
and $m=2/3$ plateaus in dependence on $J_2$ is shown in
figure~\ref{fig4}b. Obviously, both  widths shrink monotonously with
increasing of $|J_2|$.
If $J_2$ approaches the critical value   $J^{c1}_2$ we find
indications for additional plateaus, e.g., at $m=5/6$. Note, however,
that our finite-size analysis of the plateaus naturally could miss
other plateaus present in infinite systems, see, e.g., the discussion
of the ED data of the $m(h)$ curve of the standard
Shastry-Sutherland model in \cite{wir04}. Hence, the study of
the magnetization process of the considered quantum spin model needs
further attention based on alternative methods.

One might expect that the presence of these plateaus and jumps may
be linked purely to quantum effects because they are often not
observed in equivalent classical models at $T=0$
\cite{lm,kawamura,zhito,cabra}. However, for the present model the
plateau at $m=1/3$ survives in the classical limit for $J_2 < 0$ as
we will show in appendix~\ref{class}.

\subsection{Analytical results}\label{sec ar}
\subsubsection{$s=1$}
\begin{figure}
\vspace*{5cm}
\scalebox{0.85}{\includegraphics{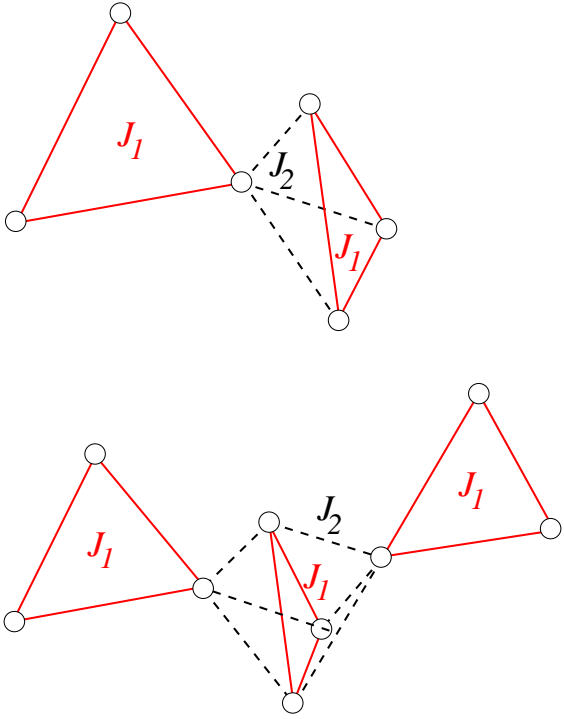}}
\caption{Two possible subsystems of the modified
Shastry-Sutherland lattice, see e.~g.~figure \ref{fig1}. The upper
one, $H_6$, consists of two coupled triangles; the lower one, $H_9$
of three triangles.
}
\label{figH69}       % Give a unique label
\end{figure}
In order to obtain analytical results about the TSPGS-region we have
adapted theorem $3$ of \cite{schmidt10} to the present situation. A
slightly more general version of this theorem is stated and proven
in appendix~\ref{app}. It yields lower bounds for the gap $\Delta E$ of the
form $\Delta E\ge f(4J)$ and the TSPGS-region in terms of properties
of simpler spin systems of which the lattice can be composed, see
figure~\ref{fig3}. These subsystems are chosen here as systems
isomorphic to $H_6$, see figure \ref{figH69}, consisting of two
neighboring triangles. For $s=1$ the gap function $x\equiv \delta_6
E=f(J),\,J\equiv\frac{J2}{J1}$ of $H_6$ is obtained as a special
case of equation (\ref{ar2a}) given below.
This yields the corresponding bounds for the
TSPGS-region $(J^{c1},J^{c2})$
\begin{equation}\label{ar3}
J^{c1}<\frac{3-\sqrt{73}}{16}\approx -0.3465<\frac{1}{4}<J^{c2}\quad
\mbox{for  }s=1\;.
\end{equation}
The function $\delta_6 E=f(J)$ according to (\ref{ar2a}) also
provides an upper bound for the gap function of the lattice, since it
represents the energy of a state orthogonal to the TSPGS, albeit not
an eigenstate of $H$. This bound is very close to the numerically
determined gap function in the case of $N=12$, see figure \ref{fig3}a,
but considerably deviates in the cases of $N=18$ and $N=24$. This
indicates that, in general, the lowest excitations of the lattice
are different from the excitations of $H_6$.\\

\subsubsection{General $s$}

\begin{figure}
\resizebox{1.0\columnwidth}{!}{
\includegraphics{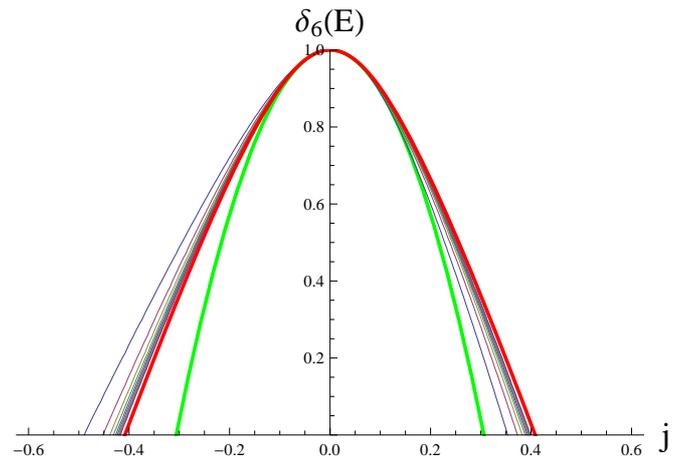}
} \caption{Lower bounds of the scaled gap function
$\delta_6 E(j),\;j=\sqrt{s(s+1)}J$ of the modified
Shastry-Sutherland spin lattice for $s=1,\ldots,10$ (thin curves)
obtained from Eq.~(\ref{ar2a}). The curves approach the asymptotic
(\ref{ar2c}) for $s\rightarrow\infty$  (thick red curve) which has a
simple quadratic approximation (\ref{ar2d}) (thick green curve).} \label{figG3b}
\end{figure}

It is possible to analytically calculate the energy of the lowest
excitations of $H_6$ for general integer $s$. The corresponding gap
$\delta_6 E=x=f(J)$ is obtained as the lowest root of the following
cubic equation
\begin{eqnarray}\nonumber
&-&(x-4) (x-2) (x-1 )
 -(x-1 ) (2x-5 ) J\\ \label{ar2a}
&+&( 1 - 3 r - x + r x)J^2 +r J^3=0
\end{eqnarray}
where we have set $r\equiv s(s+1)$. From this result one derives the
lower bound
\begin{equation}\label{ar2b}
\Delta E\ge f(4J) \quad \mbox{for general }s
\end{equation}
and a lattice of arbitrary size, see theorem $1$ in appendix~\ref{app}
adapted to the system under consideration.
The corresponding curves are shrinking in $J$-direction with
increasing $s$ and yield inner bounds for the TSPGS-region
$(J^{c1},J^{c2})$ of the form
\begin{equation}\label{ar2b1}
J^{c1} < J_{L}^{(1)}<0<J_{L}^{(2)}<J^{c2}
\;,
\end{equation}
see figure \ref{figG1} (green curves).
Upon scaling w.~r.~t.~the new variable
$j\equiv\sqrt{r}J$ the graphs of (\ref{ar2a}) asymptotically approach the curve given by
\begin{equation}\label{ar2c}
j^2=\frac{(x-4)(x-2)(x-1)}{16(x-3)}\;,
\end{equation}
with Taylor expansion
\begin{equation}\label{ar2d}
x=1-\frac{32}{3}j^2+{\mathcal O}(j^3)\;,
\end{equation}
see figure \ref{figG3b}.
Hence $J_{L}^{(i)}$ assumes for $s\to\infty$ asymptotically the form
\begin{equation}\label{ar2d1}
|J_{L}^{(i)}|\sim\frac{1}{\sqrt{6s(s+1)}},\;\;i=1,2
\;.
\end{equation}
\\

\begin{figure}
\resizebox{1.0\columnwidth}{!}{
\includegraphics{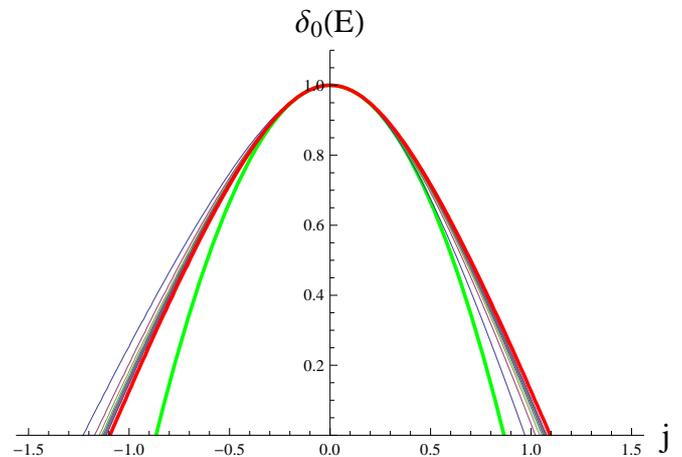}
} \caption{Upper bounds of the scaled gap function
$\delta_0 E(j),\;j=\sqrt{s(s+1)}J$ of the modified
Shastry-Sutherland spin lattice for $s=1,\ldots,10$ (thin curves)
obtained from Eq.~(\ref{ar6}). The curves approach the asymptotic
(\ref{ar7}) for $s\rightarrow\infty$  (thick red curve) which has a
simple quadratic approximation (\ref{ar8}) (thick green curve).} \label{figG3a}
\end{figure}

In order to obtain close upper bounds $g(J)$ of the gap $\Delta(E)$
in the case $N\ge 18$ we calculate
the energy of a certain (degenerate) state that involves three
triangles for arbitrary integer $s$, say, one triangle of type $I$
and two neighboring triangles of type $II$, see figure \ref{figH69}.
This state is obtained as an exact eigenstate of $H_0$, which is the
full Hamiltonian $H$, restricted to a $4^3=64-$dimensional subspace
spanned by product states of the form
\begin{equation}\label{ar4}
\phi_i\otimes\phi_j\otimes\phi_k,\quad i,j,k=0,\ldots,3\;.
\end{equation}
The $\phi_n$ live in the $(2s+1)^3$-dimensional Hilbert spaces
belonging to one of the three triangles. $\phi_0=[0,1,2]$ denotes
the TSPGS of the corresponding triangle and
\begin{equation}\label{ar5}
\phi_i\equiv \frac{\op{s}_0^{(i)}\phi_0}{||
\op{s}_0^{(i)}\phi_0||},\;i=1,2,3\;,
\end{equation}
where $\op{s}_0^{(i)}$ is the $i$-th component of the spin operator
$\op{\bf s}_0$ pertaining to the spin site number $0$, an
arbitrarily chosen spin site of the corresponding triangle. The gap
function of $H_0$ will be denoted by $x\equiv\delta_0 E=g(J)$ and
constitutes an upper bound for $\Delta(E)$.
It has the following implicit form, using $r\equiv s(s+1)$:
\begin{eqnarray}\nonumber
0&=& -12(x-3)^2(x-2)(x-1)-6J(x-3)(x-1)(4x-9)\\ \nonumber
 && +J^2
(x-3)(9-9x+4r(7x-15))+16 J^3 r (2x-5)\;.\\
&& \label{ar6}
\end{eqnarray}
Again, the function $g$ belongs to the lowest branch of (\ref{ar6}).
The corresponding curves are shrinking in $J$-direction with
increasing $s$ and yield outer bounds for the TSPGS-region
$(J^{c1},J^{c2})$ of the form
\begin{equation}\label{ar6a}
J_{U}^{(1)}<J^{c1} < 0<J^{c2}<J_{U}^{(2)}
\;,
\end{equation}
see figure \ref{figG1} (red curves).
Upon scaling w.~r.~t.~the new variable
$j\equiv\sqrt{r}J$ the graphs of (\ref{ar6})
asymptotically approach the curve given by
\begin{equation}\label{ar7}
j^2=\frac{3(x-3)(x-2)(x-1)}{7x-15}\;,
\end{equation}
with Taylor expansion
\begin{equation}\label{ar8}
x=1-\frac{4}{3}j^2+{\mathcal O}(j^3)\;,
\end{equation}
see figure \ref{figG3a}.
Hence $J_{U}^{(i)}$ assumes for $s\to\infty$ asymptotically the form
\begin{equation}\label{ar8a}
|J_{U}^{(i)}|\sim\sqrt{\frac{6}{5s(s+1)}},\;\;i=1,2
\;.
\end{equation}
\\

\begin{figure}
\resizebox{1.0\columnwidth}{!}{
\includegraphics{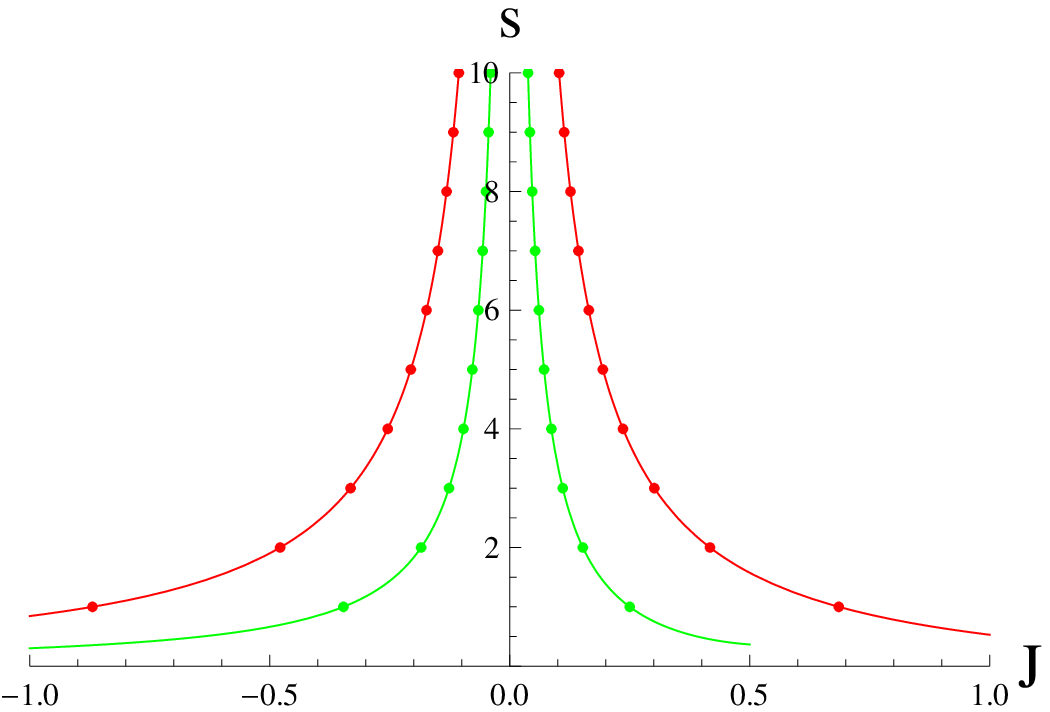}
} \caption{Exact bounds for the TSPGS-region $(J^{c1},J^{c2})$} for $s=1,\ldots,10$
of the form $J_{U}^{(1)}<J^{c1}< J_{L}^{(1)}< 0<J_{L}^{(2)}<J^{c2}<J_{U}^{(2)}$.
These are derived
from (\ref{ar2a}) (green curves, inner bounds) and (\ref{ar6}) (red curves, outer bounds).
In the classical limit $s\rightarrow\infty$
the TSPGS-region shrinks to zero according to (\ref{ar2d1}) and (\ref{ar8a}).
\label{figG1}
\end{figure}

Although these curves constitute only upper
bounds of the true gap functions, the comparison with the numerical
results for $N=18$ and $N=24$ reveals a close approximation to both
curves, see figure \ref{fig3}. This supports our conjecture that
(\ref{ar6}) indeed may serve as an analytical approximation of the
gap functions for large $N$ and arbitrary integer $s$. This would
mean that the excitations from the TSPGS can be viewed as local
excitations essentially concentrated on three neighboring triangles.
Numerically determined spin correlation functions seem to be in accordance with
this conjecture. Of course, the corresponding excited state will be
largely degenerate due to the translational symmetry of the lattice.
We expect an almost flat $\bf{k}$-dependance of the energy band
$E(\bf{k})$. This expectation is also supported by our numerical
results. We have found that the lowest excitations close to $J=0$
have the total spin quantum number $S=1$ in accordance with our
model.\\

In the case $N=12$ where we have performed numerical calculations for
$s=1$ and $s=2$ it is not possible to put a subsystem of type $H_9$  into the
lattice and the above results do not apply. However,
an analogous method can be applied to two
coupled triangles of type $H_6$ and yields an upper bound of the gap function of the form
\begin{equation}\label{ar2e}
\Delta E\le \frac{1}{12} (18 - 3 J - \sqrt{9(J-2)^2  + 96 J^2
s(s+1)}).
\end{equation}
The numerically determined gap together with the bounds (\ref{ar2b})
and  (\ref{ar2e}) is represented in figure \ref{fig3} b.\\

\section*{Acknowledgement}
The numerical calculations were performed using J.~Schulenburg's
{\it spinpack}.

\appendix

\section{Proof of the gap theorem
}
\label{app}

\begin{figure}
\resizebox{1.0\columnwidth}{!}{
\includegraphics{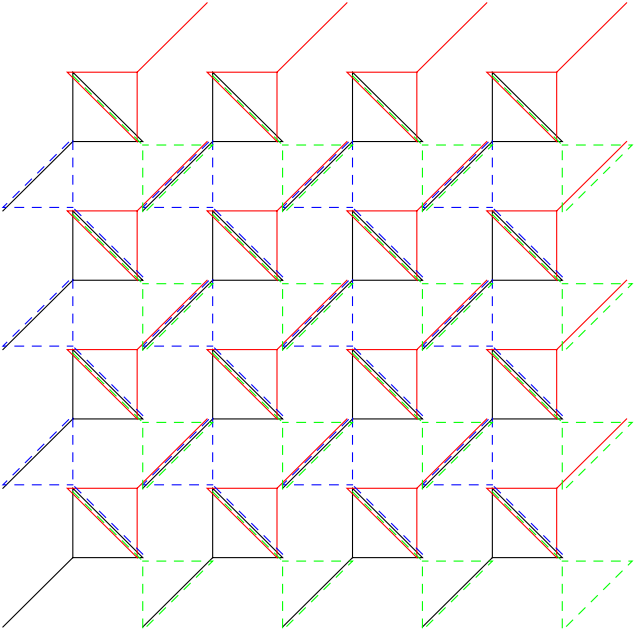}
} \caption{The Hamiltonian
of the (modified) Shastry-Sutherland
spin lattice can be written as the sum of four classes of
simpler Hamiltonians of the kind $H_6$, see figure \ref{figH69}, which are
indicated by the colors black, red, green, blue. The subsystems of each color
are isomorphic via translations $T$ of the lattice. They can be transformed into each other
by $90^\circ$ rotations $R$ about the mid-points of the empty squares. Note that there
are two types of diagonals (trimers), one with positive slope (type I) and the other
with negative one (type II). Each diagonal belongs to four subsystems of different color,
hence the coupling constant $J_1$ of the subsystems, see figure \ref{figH69}, has to be
divided by the factor $4$ in order to obtain a total Hamiltonian with the coupling constant $J_1$.} \label{figFSS}
\end{figure}

In order to prove the existence of an energy gap between the TSPGS and the first excited
state we will adapt the analogous proof given in \cite{schmidt10} to the modified Shastry-\-Sutherland
model considered in this article. The gap theorem will be formulated in a slightly more general framework.\\

For any positive integer $\ell$ let ${\mathbb Z}_{\bf \ell}$ denote
the set of integers modulo $\ell$, such that $n+\ell\equiv n$, and
\begin{equation}\label{app1}
{\mathbb Z}_{\bf L}={\mathbb Z}_{L_1}\times\ldots\times{\mathbb Z}_{L_d}
\end{equation}
a standard $d$-dimensional lattice with total size
\begin{equation}\label{app2}
{\mathcal N}=\prod_{\ell=1}^d\;L_\ell
\;.
\end{equation}
We consider an index set ${\mathcal L}$ on which the additive group ${\mathbb Z}_{\bf L}$ operates effectively,
i.~e.~without fixed points except for the neutral element. Let $c$ be the number of corresponding equivalence
classes (orbits), $c=\left|{\mathcal L}/ {\mathbb Z}_{\bf L}\right|$. Each orbit is isomorphic to ${\mathbb Z}_{\bf L}$;
if we select an index from each orbit we obtain a bijection
$\varphi:\{1,\ldots,c\}\times {\mathbb Z}_{\bf L}\longrightarrow {\mathcal L}$ which we call a ``(global) trivialization"
in analogy with the corresponding term in the theory of fibre bundles.\\
We will fix $\tau$ different  trivializations
\begin{equation}\label{app3}
\varphi_\nu:\{1,\ldots,c\}\times {\mathbb Z}_{\bf L}\longrightarrow {\mathcal L},\;\nu=1,\ldots,\tau
\;.
\end{equation}
W.~r.~t.~these trivializations the total Hilbert space can be written as a tensor product space in the
following form:
\begin{eqnarray}\label{app4a}
{\mathcal H}&=&\bigotimes_{i\in{\mathcal L}}{\mathcal H}_i
=\bigotimes_{(\ell,{\bf k})\in\{1,\ldots,c\}\times {\mathbb Z}_{\bf L}}{\mathcal H}_{\varphi_\nu(\ell,{\bf k})}\\
\label{app4b} &=&\bigotimes_{{\bf k}\in{\mathbb Z}_{\bf L}}
\bigotimes_{\ell=1,\ldots,c}{\mathcal H}_{\varphi_\nu(\ell,{\bf k})}\\ \label{app4c}
&\equiv&\bigotimes_{{\bf k}\in{\mathbb Z}_{\bf
L}}{\mathcal H}_{\nu,{\bf k}}\,,\quad\nu=1,\ldots,\tau \;.
\end{eqnarray}
We will also write
\begin{equation}\label{Snuka}
{\mathcal H}_{\nu,{\bf k}}=\bigotimes_{i\in{\mathcal S}_{\nu,{\bf k}}} {\mathcal H}_i
\mbox{ where }
{\mathcal S}_{\nu,{\bf k}} \equiv \{\varphi_\nu(\ell,{\bf k})|1\le\ell\le c\}
\;.
\end{equation}

We will explain these definitions in the case of the modified Shastry-Sutherland
model in the finite realization of figure \ref{fig1}. It is a $d=2$-dimensional lattice with
$L_1=L_2=2$, hence ${\mathcal N}=L_1 L_2=4$. The $8$ different triangles of figure \ref{fig1} correspond to the
indices $i\in{\mathcal L}$, hence ${\mathcal L}$ can be viewed as the set of triangles and be identified
with the set of numbers of their upper corners $\{16,17,\ldots,23 \}$.
Note that ${\mathcal L}$ is not isomorphic to the underlying spin lattice which has $24$ sites.
The Hilbert spaces
${\mathcal H}_i,\;i\in{\mathcal L}$ happen to be isomorphic and of the same dimension $(2s+1)^3$.
In general, it is not necessary that all ${\mathcal H}_i$ are isomorphic; e.~g.~we could have spin systems
composed of trimers and dimers.
${\mathbb Z}_{\bf L}$ operates on ${\mathcal L}$ in a natural way by means of translations. Type I triangles
$(18,19,22,23)$
cannot be transformed into type II triangles $(16,17,20,21)$ by means of translations. Hence we have, in this case,
$c=\left|{\mathcal L}/ {\mathbb Z}_{\bf L}\right|=2$. In figure \ref{figFSS} two adjacent triangles of different
types are coupled together in order to form sub-Hamiltonians of the kind $H_6$ summing over all possible translations of it.
These sub-Hamiltonians are in a $1:1$ manner characterized by trivializations of the kind considered above:
We denote the two adjacent triangles we started with by $(1,{\bf 0})$ and $(2,{\bf 0})$, and all other pairs
which are translations of these will be denoted by $(1,{\bf k})$ and $(2,{\bf k})$ where ${\bf k}$ runs through
${\mathbb Z}_{\bf L}$. Since the set ${\mathcal L}$ is exhausted by this construction we have obtained a
trivialization in the sense of (\ref{app3}), i.~e.~a bijection
$\varphi_\nu:\{1,2\}\times{\mathbb Z}_{\bf L}\longrightarrow {\mathcal L},\,\nu=1,\ldots,\tau$.
In figure \ref{figFSS} $\tau=4$ different trivializations are shown
and indicated by different colors.\\

Returning to the general case we will identify all factor spaces belonging to subsystems of the same type
by means of a certain
product basis $\{|\gamma\rangle\},\;\gamma:{\mathcal L}\longrightarrow{\mathbb N}$ in ${\mathcal H}$ such that
$i\in{\mathcal L}\mapsto\gamma(i)\in\{0,\ldots,\mbox{dim}_i\}$,
where $\mbox{dim}_i$ denotes the dimension of ${\mathcal H}_i$.
Correspondingly, the unitary translation operators $T_{\bf m},\;{\bf m}\in{\mathbb Z}_{\bf L}$ are defined by
suitable permutations of the product basis. W.~r.~t.~a trivialization $\varphi_\nu$ this definition assumes the form
\begin{equation}\label{app5}
(T_{\bf m}\,\gamma)(\varphi_\nu(\ell,{\bf n}))=\gamma(\varphi_\nu(\ell,{\bf m}+{\bf n}))
\;.
\end{equation}
The $T_{\bf m},{\bf m}\in{\mathbb Z}_{\bf L}$  form the abelian translation group ${\mathcal T}$ whose characters are of the
well-known form
\begin{eqnarray}\label{app6a}
\chi(T_{\bf m})&=&\exp \left( 2\,\pi\, i\,{\bf m}\cdot {\bf
k}/{\mathcal N} \right),\\ \label{app6b} \mbox{where} &&{\bf
m}=(m_1,\ldots,m_d),\,{\bf k}=(k_1,\ldots,k_d) \in{\mathbb Z}_{\bf
L}\; .
\end{eqnarray}
The total Hamiltonian $H$ is assumed to be a sum of sub-Hamiltonians of the form
\begin{equation}\label{app7}
H=\sum_{\nu=1}^\tau
\sum_{{\bf k}\in{\mathbb Z}_{\bf L}}
H_{\nu,{\bf k}}
\equiv
\sum_{\nu=1}^\tau
\sum_{{\bf k}\in{\mathbb Z}_{\bf L}}
T_{\bf k} H_\nu T_{\bf k}^\ast
\;,
\end{equation}
where $H_{\nu,{\bf k}}$  is defined on the ``supporting factor space" ${\mathcal H}_{\nu,{\bf k}}$,
see (\ref{app4c}), and extended
as the identity operator on the remaining factor spaces to the total space ${\mathcal H}$. Of course,
it suffices to postulate this only for $H_\nu=H_{\nu,{\bf 0}}$.

As a consequence of (\ref{app7}) we note that total Hamiltonian will commute with all translations:
\begin{equation}\label{app8}
[H,T_{\bf m}]=0 \mbox{ for all } T_{\bf m}\in{\mathcal T}
\;.
\end{equation}
Moreover, we assume the following:
\begin{ass}\label{ass1}
For all $i\in{\mathcal L}$ let $\Phi_i\in {\mathcal H}_i$ be normalized states such that
\begin{equation}\label{app9}
\Phi=\bigotimes_{i\in{\mathcal L}} \Phi_i
\end{equation}
is a ground state of $H_\nu$ for all $\nu=1,\ldots,\tau$ which is unique on the factor space  ${\mathcal H}_{\nu,{\bf 0}}$.
We set
\begin{equation}\label{app10}
H_\nu\Phi=E_\nu^{(0)}\,\Phi
\end{equation}
and denote by
\begin{equation}\label{app11}
E_\nu^{(1)}=E_\nu^{(0)}+\delta_\nu,\;\delta_\nu > 0
\end{equation}
the next-lowest energy eigenvalue of $H_\nu$.
Moreover, $\Phi$
is assumed to be invariant under translations,
\begin{equation}\label{app12}
T_{\bf m}\Phi=\Phi\mbox{ for all  }T_{\bf m}\in{\mathcal T}
\;.
\end{equation}
\end{ass}
Then the gap theorem can be formulated as follows.
\begin{theorem}
\label{T1}
Under the preceding definitions and assumptions
$\Phi$ will be the unique
ground state of $H$ with eigenvalue $\tilde{E}_0={\mathcal N}\,\sum_{\nu=1}^\tau E_\nu^{(0)}$ and the next-lowest
eigenvalue of $H$ satisfies $\tilde{E}_1\ge \tilde{E}_0+\sum_{\nu=1}^\tau\,\delta_\nu$.
\end{theorem}
The existence of a gap follows since $\sum_{\nu=1}^\tau\,\delta_\nu>0$ is independent of the size ${\mathcal N}$
of the lattice.
In the special but important case where all $H_\nu,\;\nu=1,\ldots,\tau$ are unitarily equivalent, we write
$E^{(1)}=E^{(0)}+\delta$ and conclude
$\tilde{E}_0={\mathcal N}\,\tau\,E^{(0)}$ and $\tilde{E}_1\ge \tilde{E}_0+\tau\,\delta$.\\

\noindent {\bf Proof of theorem \ref{T1}}:
The first claim (except uniqueness) follows immediately from assumption \ref{ass1} and the fact that,
due to \ref{app12},
$\Phi$ is also a ground state of
all $H_{\nu,{\bf k}}$ with the same eigenvalue $E_\nu^{(0)},\;{\bf k}\in{\mathbb Z}_{\bf L}$
and ${\mathcal N}\,\sum_{\nu=1}^c E_\nu^{(0)}$ being an obvious lower bound of $H$.
\\
Let
$\Psi\in\mathcal{H}$ be the eigenvector of $H$ belonging to the
next-lowest eigenvalues $\tilde{E}_1\ge\tilde{E}_0$.
We first note that $\Psi\perp\Phi$ follows
in the case $\tilde{E}_1>\tilde{E}_0$ and can be arranged in the case
$\tilde{E}_1=\tilde{E}_0$ (which we cannot exclude from the outset) by choice of $\Psi$.
Moreover, due to (\ref{app8}) we may choose $\Psi$ to be a common eigenvector of all translations,
\begin{equation}\label{app13}
T_{\bf m}\,\Psi = \chi(T_{\bf m})\,\Psi\mbox{ for all   }T_{\bf m}\in{\mathcal T}
\;,
\end{equation}
where $\chi(T_m)$ is of the form (\ref{app6a}).

Our aim is to show $\tilde{E}_1\ge \tilde{E}_0+\sum_{\nu=1}^c\,\delta_\nu$.
Let $|{\alpha}\rangle,\alpha=0,1,2,\ldots$ denote the eigenbasis of $H_{\nu,{\bf k}}$ in
${\mathcal H}_{\nu,{\bf k}}$.
Further we arrange the eigenbasis such that
$|{0}\rangle = \bigotimes_{i\in{\mathcal S}_{\nu,{\bf k}}}\,\Phi_i$ holds, see (\ref{Snuka}).
The corresponding eigenvalues of  $H_{\nu,{\bf k}}$ are denoted by $E_\nu^{(\alpha)}$,
in accordance to the notation $E_\nu^{(0)}$ and $E_\nu^{(1)}$ introduced above.
$|{\alpha,K}\rangle$ denotes a corresponding product basis
in $\mathcal{H}$, where $K$ stands for some multi-index of quantum numbers.
Moreover, we consider the reduced density operator
$W_\Psi^{\nu,{\bf k}}$ in ${\mathcal H}_{\nu,{\bf k}}$
defined by the partial trace
\begin{equation}\label{app14}
\langle\alpha|W_\Psi^{\nu,{\bf k}}|\beta\rangle = \sum_K \langle\alpha,K|\Psi\rangle\langle\Psi|\beta,K\rangle
\;.
\end{equation}

Then we conclude
\begin{eqnarray}\label{app15a}
\tilde{E}_1&=&\langle \Psi|H|\Psi\rangle = \sum_{\nu,{\bf k}} \langle \Psi|H_{\nu,{\bf k}}|\Psi\rangle\\
\label{app15b}
&=& \sum_{\nu,{\bf k}} \mbox{Tr}\left( H_{\nu,{\bf k}}\,W_\Psi^{\nu,{\bf k}}\right)\\
\label{app15c}
&=&\sum_{\nu,{\bf k},\alpha} \mbox{Tr}\left( E_\nu^{(\alpha)} |\alpha\rangle\langle\alpha|W_\Psi^{\nu,{\bf k}}\right)\\
\label{app15d}
&=&\sum_{\nu,{\bf k},\alpha}E_\nu^{(\alpha)}\langle\alpha|W_\Psi^{\nu,{\bf k}}|\alpha\rangle\\
\label{app15e}
&=&\sum_{\nu,{\bf k}}\left(
E_\nu^{(0)} \langle 0|W_\Psi^{\nu,{\bf k}}|0\rangle+\sum_{\alpha>0}
E_\nu^{(\alpha)}\langle\alpha|W_\Psi^{\nu,{\bf k}}|\alpha\rangle
\right)\\ \nonumber
&\ge&
\sum_{\nu,{\bf k}}\left(
E_\nu^{(0)} \langle 0|W_\Psi^{\nu,{\bf k}}|0\rangle+(E_\nu^{(0)}+
\delta_\nu)\sum_{\alpha>0}\langle\alpha|W_\Psi^{\nu,{\bf k}}|\alpha\rangle
\right)
\;.\\
&&\label{app15f}
\end{eqnarray}

\begin{lemma}
\begin{equation}\nonumber
\langle 0|W_\Psi^{\nu,{\bf k}}|0\rangle \le 1-\frac{1}{\mathcal N}\;.
\end{equation}
\label{lemma1}
\end{lemma}
\noindent {\bf Proof of lemma \ref{lemma1}}:
It suffices to consider the case ${\bf k}={\bf 0}$. We again consider the product
basis $\{|\gamma\rangle\}$ introduced above and write the $c$ quantum numbers
$\gamma(\varphi_\nu(\ell,{\bf 0})),\;\ell=1,\ldots,c,$ at the first $c$ places of the
string $|\gamma\rangle=$ \\
$|n_1,n_2,\ldots,n_c,\ldots\rangle$. It follows that the ground state
in ${\mathcal H}_{\nu,{\bf 0}}$ is denoted by a ket $|0,0,\ldots,0\rangle$ consisting of $c$ zeroes.

We conclude
\begin{equation}\label{app16}
\langle 0,\ldots,0|W_\Psi^{\nu,{\bf 0}}|0,\ldots,0\rangle = \sum_K |\langle\Psi|0,\ldots,0,K\rangle|^2\equiv s_0
\end{equation}
and
\begin{eqnarray} \label{app17a}
1&=&\mbox{Tr}W_\Psi^{\nu,{\bf 0}}\\ \label{app17b}
&=&\sum_{n_{c+1},n_{c+2},\ldots} |\langle\Psi|0,\ldots,0,n_{c+1},n_{c+2},\ldots\rangle|^2\\
\label{app17c}
&&+ \sum_{n_1,n_2,\ldots}|\langle\Psi|n_1,n_2,\ldots\rangle|^2 \\ \label{app17d}
&\equiv& s_0+s_1
\;.
\end{eqnarray}
The first sum $s_0$ in (\ref{app17a}) runs through all sequences\\ $0,\ldots,0,n_{c+1},n_{c+2},\ldots$
excluding the value $n_{c+1}=n_{c+2}=\ldots=0$, since $\langle\Psi|\Phi\rangle=0$.
Equivalently, we will say that it runs through all states
$\psi=|0,\ldots,0,n_{c+1},n_{c+2},\ldots\rangle\in\mathcal{B}_0$.
The second sum $s_1$ in (\ref{app17b}) runs through all sequences $n_1,n_2,\ldots$
except those with $n_1=n_2=\ldots=n_c=0$, or,
equivalently, through all states $\psi=|n_1,n_2,\ldots\rangle\in\mathcal{B}_1$.
Thus the total sum in (\ref{app17a},\ref{app17b}) runs through an orthonormal basis
$\mathcal{B}=\mathcal{B}_0\cup \mathcal{B}_1$ of
$\mathcal{H}'\equiv \{\psi\in\mathcal{H}|\langle\psi|\Phi\rangle=0\}$.\\
We consider on $\mathcal{B}$ the equivalence relation
$\psi_1 \sim \psi_2 \Leftrightarrow \psi_1=T_{\bf m}\,\psi_2$ for some $T_{\bf m}\in{\mathcal T},$
and denote by $\Lambda=\mathcal{B}/_\sim$ the corresponding set of equivalence classes or ``orbits".
Due to (\ref{app13}) all states $\psi$ in the same orbit $\lambda$ yield the same value
\begin{equation}\label{app18}
t_\lambda\equiv |\langle\Psi|\psi\rangle|^2=|\langle\Psi|T_{\bf m}\,\psi\rangle|^2,\;T_{\bf m}\in{\mathcal T}
\;.
\end{equation}
For each orbit $\lambda\in\Lambda$ let $N_\lambda\equiv|\lambda|$ denote its length.
For most orbits we have $N_\lambda={\mathcal N}$, but in general $N_\lambda$ will be a divisor of ${\mathcal N}$.
For example, if $d=1$, $L_1={\mathcal N}=6$ and $|1,2,3,1,2,3\rangle\in\lambda$ then $N_\lambda=3$.
We define $N_\lambda^{(k)}\equiv|\lambda\cap\mathcal{B}_k|,\;k=0,1,$ and obtain the
following equations:
\begin{eqnarray}\label{app19}
N_\lambda&=&N_\lambda^{(0)}+N_\lambda^{(1)}\;,\\
s_0&=& \sum_{\lambda\in\Lambda}t_\lambda\,N_\lambda^{(0)}\;,\\
s_1&=& \sum_{\lambda\in\Lambda}t_\lambda\,N_\lambda^{(1)}
\;.
\end{eqnarray}
Let $\psi=|n_1,n_2,n_3,n_4,\ldots\rangle\in\lambda$. Note that at least one $n_j,\;1\le j\le c{\mathcal N}$ must
be non-zero since $\psi\neq\Phi=|0,0,\ldots,0\rangle$.
Hence at least one translation of $\psi$ belongs to $\mathcal{B}_1$, namely that where $j$ is shifted to one of the
first $c$ places.
To show this in detail we write $n_j=\gamma(\varphi_\nu(\ell,{\bf k})),\;1\le\ell\le c,\,{\bf k}\in{\mathbb Z}_{\bf L}$.
It follows that $T_{-{\bf k}}\psi=|m_1,m_2,m_3,m_4,\ldots\rangle\in\lambda$ and $m_\ell=n_j\neq 0$.
Thus $N_\lambda^{(1)}\ge 1$ and hence $N_\lambda^{(0)}\le N_\lambda-1\le {\mathcal N}-1$
which for $\sum_{\lambda\in\Lambda}t_\lambda >0$ implies
\begin{equation}\label{app20}
\frac{s_1}{s_0}\ge\frac{1}{{\mathcal N}-1}
\;.
\end{equation}
$\sum_{\lambda\in\Lambda}t_\lambda=0$ is impossible since it would imply that $t_\lambda=0$ for all $\lambda\in\Lambda$
and hence $1=s_0+s_1=0$.\\
From (\ref{app20}) we infer
\begin{equation}\label{app21}
\frac{1}{s_0}=\frac{s_0+s_1}{s_0}=1+\frac{s_1}{s_0}\ge 1+\frac{1}{{\mathcal N}-1}=\frac{{\mathcal N}}{{\mathcal N}-1}
\end{equation}
and
\begin{eqnarray}\label{app22a}
s_0&\le& \frac{{\mathcal N}-1}{{\mathcal N}}=1-\frac{1}{{\mathcal N}}\;,\\ \label{app22b}
s_1&\ge& \frac{1}{{\mathcal N}}
\;,
\end{eqnarray}
which concludes the proof of the lemma.
\qed\\

\noindent To complete the proof of theorem \ref{T1} we use again the eigenbasis
of $H_{\nu,{\bf k}}$ and write
\begin{equation}\label{app23}
1=\mbox{Tr }W_\Psi^{\nu,{\bf k}}=\langle 0|W_\Psi^{\nu,{\bf k}}|0\rangle+
\sum_{\alpha=1,2,\ldots}\langle\alpha|W_\Psi^{\nu,{\bf k}}|\alpha\rangle
=s_0+s_1
\;.
\end{equation}
Then we rewrite (\ref{app15f}) as
\begin{eqnarray}\label{app24a}
\tilde{E}_1&\ge& \sum_{\nu,{\bf k}} \left(
E_\nu^{(0)}\,s_0+(E_\nu^{(0)}+\delta_\nu)\,s_1
\right)\\ \label{app24b}
&=&
\sum_{\nu,{\bf k}}(E_\nu^{(0)}+\delta_\nu\,s_1)\\ \label{app24c}
&=& \tilde{E}_0+{\mathcal N}\,\sum_{\nu}\delta_\nu\,s_1\ge \tilde{E}_0+
{\mathcal N}\,\frac{1}{{\mathcal N}}\sum_{\nu}\delta_\nu\\ \label{app24d}
&=&\tilde{E}_0+\sum_{\nu=1}^\tau\delta_\nu
\;,
\end{eqnarray}
where we have used (\ref{app22b}) which is equivalent to lemma \ref{lemma1}.
\qed\\

In order to apply the gap theorem to the modified Shastry-Sutherland lattice we write for the
energies of the subsystems  $H_{\nu,{\bf k}}$
\begin{equation}\label{app25}
E^{(1)}=E^{(0)}+\delta({\scriptstyle \frac{1}{4}},J_2)
\;.
\end{equation}
This has the consequence that the total Hamiltonian will correspond to the coupling constants $J_1=1$ and $J_2$
since each triangle is contained in four different subsystems, see figure \ref{figFSS}.
Since $\delta$ is a homogeneous function of $J_1,\,J_2$,
i.~e.~$\delta(\alpha J_1,\alpha J_2)=\alpha\,\delta(J_1,J_2)$, we may write
$\delta(\frac{1}{4},J_2)=\frac{1}{4}\,\delta(1,4 J_2)$. Hence the gap theorem implies (note that $\tau=4$)
\begin{equation}\label{app26}
\tilde{E}_0\ge\tilde{E}_1+4\,\frac{1}{4}\,\delta(1,4J_2)=\tilde{E}_1+\delta(1,4J_2)
\;.
\end{equation}
Thus the lower bound of the gap is simply obtained by shrinking the graph of the
gap function $\delta(1,J_2)$ of $H_6$ into $J_2$-direction by a factor $4$.

\section{Classical ground states
}
\label{class}

\begin{figure}
\resizebox{1.0\columnwidth}{!}{
\includegraphics{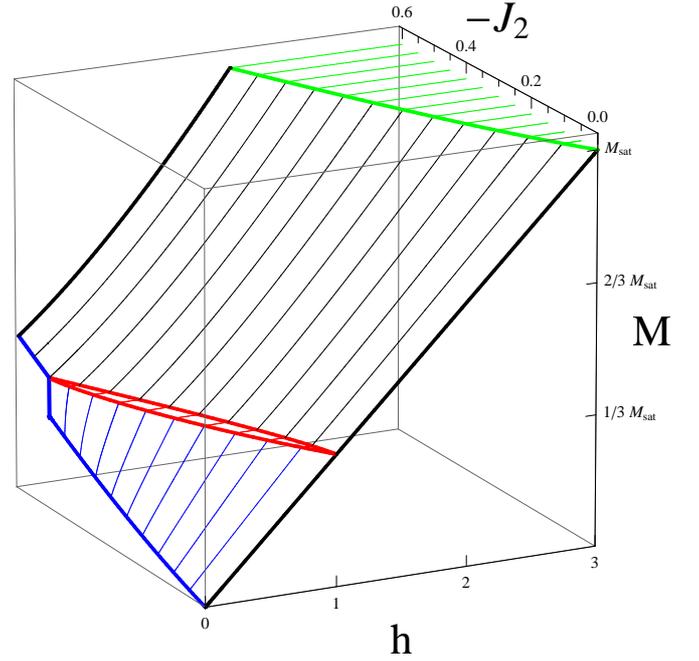}
} \caption{ Magnetization curves for the classical modified Shastry-Sutherland lattice for different values of
$J=-J_2$ between $0$ and $0.6$. The magnetization plateau at $M=\frac{1}{3}M_{\mbox{\scriptsize sat}}$ (indicated by red color)
separates states of phase II (blue curves) from those of phase III (black curves). Full magnetization $M=M_{\mbox{\scriptsize sat}}$
is reached in the region indicated by green color. The phase boundaries can be analytically calculated,
see (\ref{class7}) - (\ref{class11}).}
\label{figmc}
\end{figure}

\begin{figure}
\resizebox{1.3\columnwidth}{!}{
\includegraphics{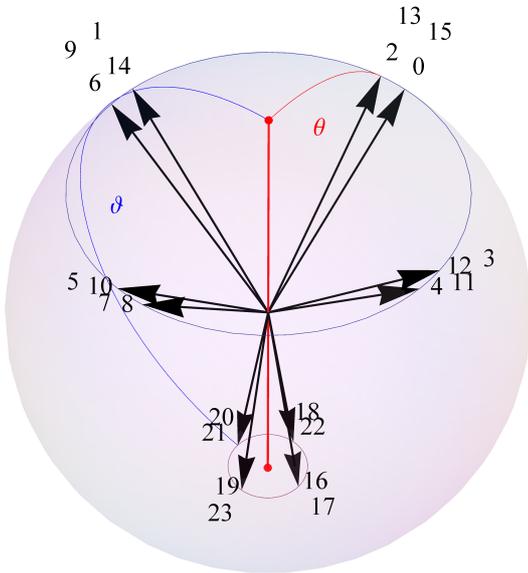}
} \caption{A typical classical ground state of phase II according to (\ref{class2})-(\ref{class5}).
The numbers $\mu$ which are attached to the spin vectors ${\bf s}_\mu$ refer to the spin sites of figure
\ref{fig1}. All spin vectors of the in-plane spins lie on a cone with opening angle $\theta$, whereas the out-of-plane
spins have a polar angle $\vartheta$.}
\label{figcs}
\end{figure}

It follows from the general theory \cite{schmidt10} as well as from our special results
(\ref{ar2c}) and (\ref{ar7}) that the classical modified Shastry-Sutherland model
possesses no TSPGS's except for $J_2=0$. Nevertheless it is possible to analytically obtain the classical
ground states for $J_2\neq 0$ and arbitrary magnetic field $h$  and from these the magnetization curves.
Typically in the classical limit $s\rightarrow\infty$ the magnetization curves
at $T=0$ are smooth and do
not exhibit plateaus or jumps \cite{lm,kawamura,zhito,cabra}.
An exception to this rule is, e.~g., reported in \cite{schroeder05}. Hence
it is remarkable that the classical modified Shastry-Sutherland model possesses a plateau at a magnetization
of $M=\frac{1}{3}M_{\mbox{\scriptsize sat}}$ and a jump at $h=0,\;J_2=-1/2$, as will be shown in the sequel.\\

For sake of simplicity we assume a quadratic square lattice of $L\times L$ squares, where $L\ge 4$ is some multiple of $4$.
It hence contains $\ell=\frac{L^2}{2}$ triangles $\Delta_\mu,\;\mu=1,\ldots,\ell$.
Let ${\bf s}_{\mu,0},\,{\bf s}_{\mu,1},\,
{\bf s}_{\mu,2}$ denote the three (unit) spin vectors corresponding to $\Delta_\mu$ such that ${\bf s}_{\mu,0}$ corresponds
to the ``out-of-plane" spin site, see figure \ref{fig1}, and
\begin{equation}\label{class1}
{\bf S}_\mu={\bf s}_{\mu,0}+{\bf s}_{\mu,1}+
{\bf s}_{\mu,2},\quad \mu=1,\ldots,\ell
\end{equation}
denote its total spin. $\Delta_\mu$ is uniformly coupled to two
adjacent sites which belong to neighboring triangles with strength
$J_2$. As usual, we write the Zeeman term in the Hamiltonian as
$-h\,M\equiv -h\,\left( \sum_{\mu=1}^\ell {\bf S}_\mu\right)_3$ where
$h$ is the strength of the (dimensionsless) magnetic field. We will
confine ourselves to the ferromagnetic case $J_2<0$, which shows the
most interesting features.
In the AF case $J_2>0$ the magnetization curves are almost linear until they reach the saturation domain.\\
As one can see in figure \ref{figmc} there are, besides the fully aligned state with $M=M_{\mbox{\scriptsize sat}}=3\ell$,
exactly three different phases.
Phase I which forms the magnetization plateau at $M=\frac{1}{3}M_{\mbox{\scriptsize sat}}=\ell$ is given by the $uud$-state,
i.~e.~in each triangle $\Delta_\mu$ the two in-plane spins point into the direction
of the magnetic field (``up") and the off-plane spin in the opposite direction
(``down"). \\
The two other phases II and III have spin vectors of the form
\begin{equation}\label{class2}
{\bf s}_{\mu\,i}=\Vektor{\sin\theta\,\cos \varphi_{\mu\,i}}{\sin\theta\,\sin \varphi_{\mu\,i}}{\cos\theta},\quad \mu=1,\ldots,\ell,\;i=1,2
\end{equation}
and
\begin{equation}\label{class3}
{\bf s}_{\mu\,0}=\Vektor{\sin\vartheta\,\cos \varphi_{\mu\,0}}{\sin\vartheta\,\sin \varphi_{\mu\,0}}{\cos\vartheta},\quad \mu=1,\ldots,\ell\;.
\end{equation}
Ground states of phase II live in the domain $0<M<\frac{1}{3}M_{\mbox{\scriptsize sat}}$ and are visualized in figure \ref{figcs}.
Their azimuthal angles $\varphi_{\mu\,i},\,i=1,2$ assume $8$ different values
\begin{equation}\label{class4}
\varphi_{\mu\,i}=n\frac{\pi}{2}\pm \phi,\;n=0,\ldots,3
\end{equation}
which depend on a parameter $\phi$, whereas
\begin{equation}\label{class5}
\varphi_{\mu\,0}=n\frac{\pi}{2},\;n=0,\ldots,3
\;.
\end{equation}
Upon a translation from, say, the triangle $(0,7,16)$ to $(4,9,19)$, see figure \ref{fig1},
the azimuthal angles (\ref{class4}) and (\ref{class5}) are shifted by an amount of $-\frac{\pi}{2}$.
Hence the states of phase II are characterized by a wave number of $k=\frac{\pi}{2}$ (the minus sign does not matter).
If the magnetic field $h$ approaches the left hand boundary of the $M=\frac{1}{3}M_{\mbox{\scriptsize sat}}$ plateau,
the states of phase II become the $uud$-state.
Phase III is confined to a magnetization $M$ satisfying $\frac{1}{3}M_{\mbox{\scriptsize sat}}<M<M_{\mbox{\scriptsize sat}}$.
The corresponding states have a wave number $k=\pi$, since their azimuthal angles satisfy $\varphi_{\mu\,i}\in\{0,\pi\}$
for $\mu=1,\ldots,\ell$ and $i=1,2,3$. More precisely, the spins in figure \ref{fig1} with the numbers
$0,2,5,7,8,10,13,15;18,19,22,23$ have an azimuthal angle of $\varphi=0$, and the remaining spins of $\varphi=\pi$.\\
It is obvious how to generalize the states of phase I, II, III to infinite modified Shastry-Sutherland lattices by periodic
continuation. The (semi-)analytical treatment of these states can be based on the ground state equation, see \cite{schmidt03},
\begin{equation}\label{class6}
{\bf j}_\mu\equiv\sum_\nu J_{\mu\nu} {\bf s}_\nu-\Vektor{0}{0}{h}=\kappa_\mu\,{\bf s}_\mu,\quad\mu=1,\ldots,N
\;.
\end{equation}
Note that here Greek indices $\mu,\nu=1,\ldots,N$ do not number triangles but all spins of a finite spin lattices.
The $\kappa_\mu$ denote Lagrange parameters due to the constraints ${\bf s}_\mu\cdot{\bf s}_\mu=1$.
They can be eliminated by writing ${\bf j}_\mu\times{\bf s}_\mu={\bf 0},\;\mu=1,\ldots,N$.
These equations, together with the corresponding periodicity properties of phase II or III states,
are sufficient to determine the unknowns $\theta,\,\vartheta$ and, for phase II states, $\phi$, as solutions
of certain algebraic equations that contain $J_2$ and $h$ as parameters.
We have calculated the magnetization curves of figure \ref{figmc} by means
of numerical solutions of these equations and checked the results by a
direct numerical calculation of the ground states for different $h$ and $J$. For $h=0$ we found a large
degenerate set of ground states with total spin $S$ varying between $0$ and some $S_{\mbox{\scriptsize max}}$.
In the limit $h\rightarrow 0$ only the states with $S=S_{\mbox{\scriptsize max}}$ are obtained as ground states
with finite magnetization $M=S_{\mbox{\scriptsize max}}$.

Actually, the phase boundaries, displayed in figure \ref{figmc} by thick lines, can be given in closed form.
We will write $J=-J_2$ and remind the reader of $J_1=1$ and $\ell=\frac{L^2}{2}=\frac{1}{3}M_{\mbox{\scriptsize sat}}$.
The curve for $h\rightarrow 0$ (thick blue curve in figure \ref{figmc}) consists of three parts, the phase II part
\begin{equation}\label{class7}
M_{II}(J)=\ell\,J\,\sqrt{1+2J},\quad 0\le J\le \frac{1}{2}
\;,
\end{equation}
the phase III part
\begin{equation}\label{class8}
M_{III}(J)=\frac{\ell}{2}\,\sqrt{J(1+2J)(3+2J)},\quad\frac{1}{2}\le J < \frac{3}{2}
\;,
\end{equation}
and a jump at $J=\frac{1}{2}$ from $M_1=\ell\frac{\sqrt{2}}{2}$ to $M_2=\ell$.\\
The saturation field $h_{\mbox{\scriptsize sat}}(J)$ (thick green curve in figure \ref{figmc}) is given by
\begin{equation}\label{class9}
h_{\mbox{\scriptsize sat}}(J)=\frac{1}{2}\left(
3-11J+\sqrt{3(3+10J+19J^2)}
\right),\, 0\le J \le \frac{3}{2}
\;.
\end{equation}
The value $h_{\mbox{\scriptsize sat}}(0)=3$
is part of the linear magnetization curve for $J_2=0$ (thick black line in figure \ref{figmc})
$M(h)= \ell\, h$ which corresponds to the magnetization of a uniform AF triangle.
\\
The two boundaries (\ref{class8}) and (\ref{class9}) meet at the point $h=0,\,J=\frac{3}{2},\,M=M_{\mbox{\scriptsize sat}}=3\ell$.\\

Finally, the plateau at $M=\frac{1}{3}M_{\mbox{\scriptsize sat}}$ (bounded by thick red curves in figure \ref{figmc})
is given by the inequalities
$h_{II}(J)\le h \le h_{III}(J)$ where $h_{II}(J)$ is the lower positive root of
\begin{equation}\label{class10}
h(h-1)^2+4(h-1)(2h-1)J+(19h-16)J^2+16 J^3=0
\end{equation}
in the interval $0\le J \le \frac{1}{2}$ and
\begin{equation}\label{class11}
h_{III}(J)=\frac{1}{2}\left(
1-9J+\sqrt{1+14J+17J^2}
\right),\quad 0\le J \le \frac{1}{2}
\;.
\end{equation}
Obviously, the graphs of $h_{II}$ and $h_{III}$ intersect at the two points $J=0,\,h=1$
and $J=\frac{1}{2},\,h=0$.

%
% BibTeX users please use
% \bfbliographystyle{}
% \bfbliography{}
%
% Non-BibTeX users please use

\end{document}